\documentclass[aip,rsi,twocolumn]{revtex4}
\usepackage{amsmath}
\usepackage{amsfonts}
\usepackage{amssymb}
\usepackage{dcolumn}
\usepackage{bm}
\usepackage{graphicx}
\usepackage{epstopdf}
\usepackage{color}
\def \ie {{\it i.e.} }
\def \In {$^{115}$In }
\def \P {$^{31}$P }

\begin{document}

\title{Application of Surface Coil for Nuclear Magnetic Resonance  Studies of Semi-conducting Thin Films} 

\author{Wencong Liu, Lu Lu and V. F. Mitrovi{\'c}}
\affiliation{Physics Department, Brown University, Providence, RI 02912 USA}%
\date{\today}  
\begin{abstract}
We conduct a comprehensive set of tests of performance of surface coils used for nuclear magnetic resonance (NMR)   study of quasi 2-dimensional samples. We report ${^{115} \rm{In}}$ and ${^{31} \rm{P}}$  NMR measurements on InP,   semi-conducting thin substrate samples.  Surface coils of both zig-zag meander-line  and concentric spiral geometries were used. We compare reception sensitivity and signal-to-noise ratio (SNR) of NMR  signal obtained by using  surface-type coils to that obtained by standard  solenoid-type coils. As expected, we find that surface-type coils  provide better sensitivity for NMR study of thin films samples. Moreover, we compare the reception sensitivity of different types of the surface coils. We identify the optimal geometry of the surface coils for a given application and/or direction of the applied magnetic field.


\end{abstract}

\pacs{Valid PACS appear here}
\maketitle

\section{\label{sec:level1}INTRODUCTION}

The nuclear magnetic resonance (NMR) technique is a very powerful scientific tool, both in medical imaging and basic science. 
More precisely, NMR is a bulk microscopic probe of magnetism that can be used to effectively determine   spatial variations of local magnetic properties in matter. Therefore,   additionally to being a valuable tool for studying ordered magnetic states,   it can be used to access real space features in spatially inhomogeneous states and those characterized by short range magnetic order. Furthermore, the element site-specific nature of our probe permits the separation of itinerant from local  electronic properties. The applicability of NMR over a broad range of magnetic fields up to 45 T allows an unparalleled exploration of phase diagrams. However, the problem is that detection of the NMR signal typically requires  an order of $10^{20}$ spins. 
For this reason, the application of NMR to study properties of   nano-structures, 2D materials, and devices is precluded. 

\begin{figure}[b]
\begin{minipage}{0.98\hsize}
\centerline{\includegraphics[scale=0.35]{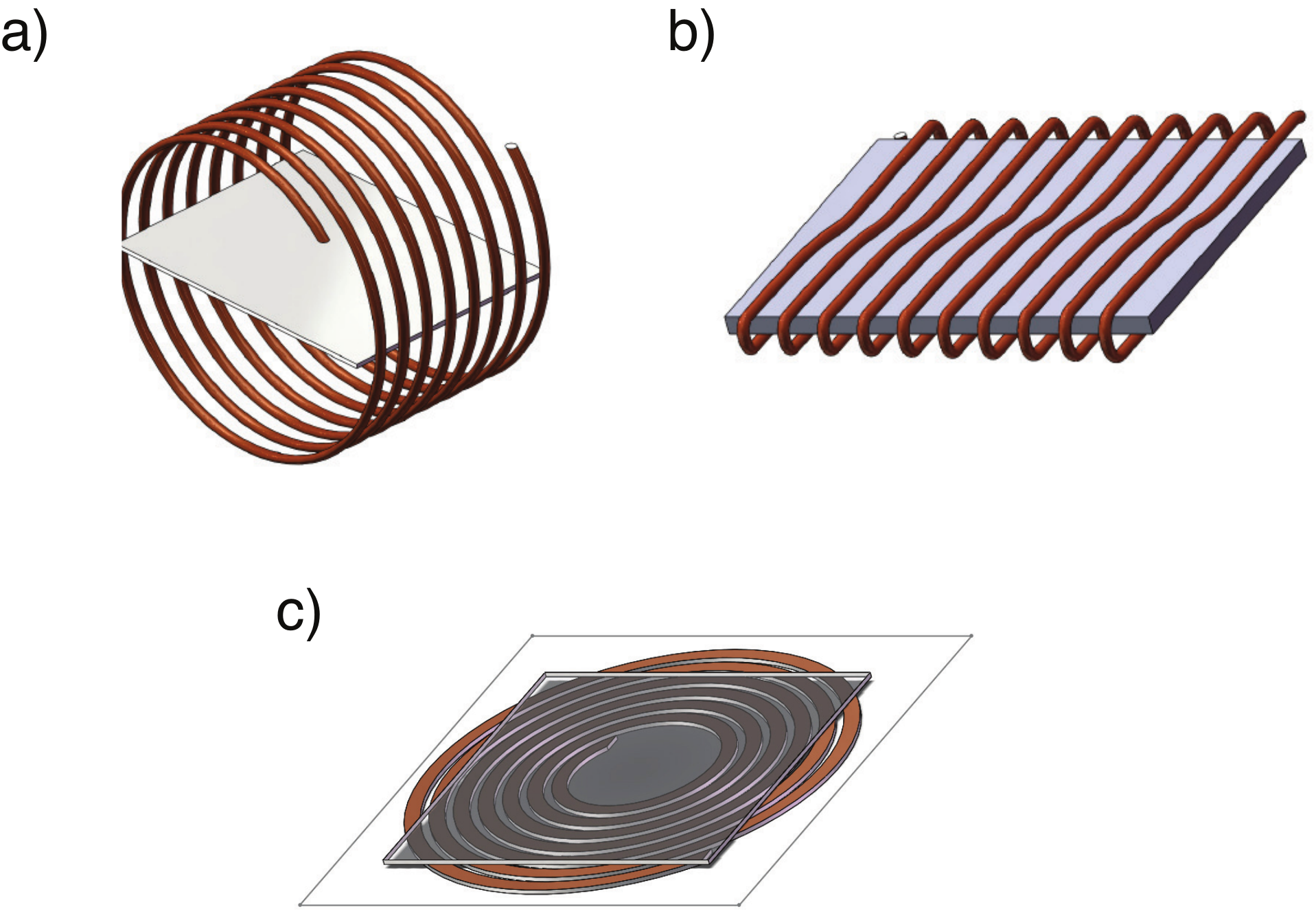}} 
\begin{minipage}{1\hsize}
\caption[]{\label{Fig1Coils}{Sketches of different type of coils typically used in an NMR experiment:} {\bf a)} solenoid, {\bf b)} flat solenoid, and  {\bf c)} spiral surface coil. Flat  gray boxes  illustrate thin samples used in our study.}   
\vspace*{-0.2cm}
\end{minipage}
\end{minipage}
\end{figure}

Inherently low sensitivity of NMR can be improved by enhancing the difference of Boltzmann population  of the nuclear Zeeman energy levels by either lowering temperature and/or increasing the applied magnetic field.  However,  we often require knowledge of physical properties of matter as a function of temperature and/or applied field. Thus, sensitivity has to be improved by varying some other parameter. 
The signal-to-noise ratio (SNR) in an NMR experiment is also controlled by the filling fraction, which is the  ratio of the volume of the effective radio frequency (RF) field to the sample volume \cite{AbragamBook}.  Consequently, by increasing the filling factor of the solenoid coil, the sensitivity  can be improved, as illustrated in \mbox{Fig. \ref{Fig1Coils}}. Nonetheless, this strategy does not work for volume limited samples and/or 2D thin-films. In such cases the application of surface  micro-coils can be beneficial \cite{Buess91, Eroglu03, Ackerman80}. This is because the effective RF field strength rapidly decays   away from the surface of the coil  \cite{Buess91}.  Since  the effective RF field  is confined to a region adjacent to the coil,   the filling factor, and thus SNR, is maximized for   2D-like samples or for probing sizable surface area to a limited depth. 
In this paper, we investigate the performance of surface coils for NMR studies of  quasi 2D materials. We compare the reception sensitivity of different geometries of surface coils and identify one which is optimal for a given application.
We emphasize that in addition to  the gain in sensitivity for such samples the application of the surface coils offers ready access to the sample. This is important if other external parameters, such as bias gate voltages and/or applied strain are to be varied.

  %
%

All surface coils produce an RF field whose magnitude decays away from the coil in the direction perpendicular to its plane. 
The inhomogeneity of the RF excitation field is  one of the major challenges in the application of surface coils to imaging and high resolution NMR spectroscopy \cite{BlumichBook}. However,   for the following reasons, this is not an issue in the NMR study of  fundamental properties of quantum materials.  Often intrinsic magnetic and/or electronic inhomogeneities and textures far exceed   the inhomogeneity of the excitation field generated by the surface coil \cite{Lu17, MitrovicLSCO}. The penetration of the RF field is spatially varying due to electronic and superconducting currents shielding effect in  metallic and superconducting samples, respectively \cite{Koutroulakis08}. 
As far as applications to condensed matter physics examined here, the main advantage for using surface coils is increased sensitivity for resonance studies of devices and quasi 2D materials,  tuning of  the probing  depth of sizable surface area of the sample, and ready access to the sample.

The paper is organized as follows. We  discuss basic principles  and assumptions necessary for the intuitive understanding of the NMR performance of different surface coils in \mbox{section \ref{theory}}. In \mbox{section \ref{exp}} we describe experimental details about our NMR set-up, coil fabrication, and the samples.  Measurements of the NMR spectra acquired by separate surface coils and for the different orientations of the magnetic field are presented  in \mbox{section \ref{res}}.

  %
%
\begin{figure}[t]
  \vspace*{-0.3cm}
\begin{minipage}{0.98\hsize}
\centerline{\includegraphics[scale=0.53]{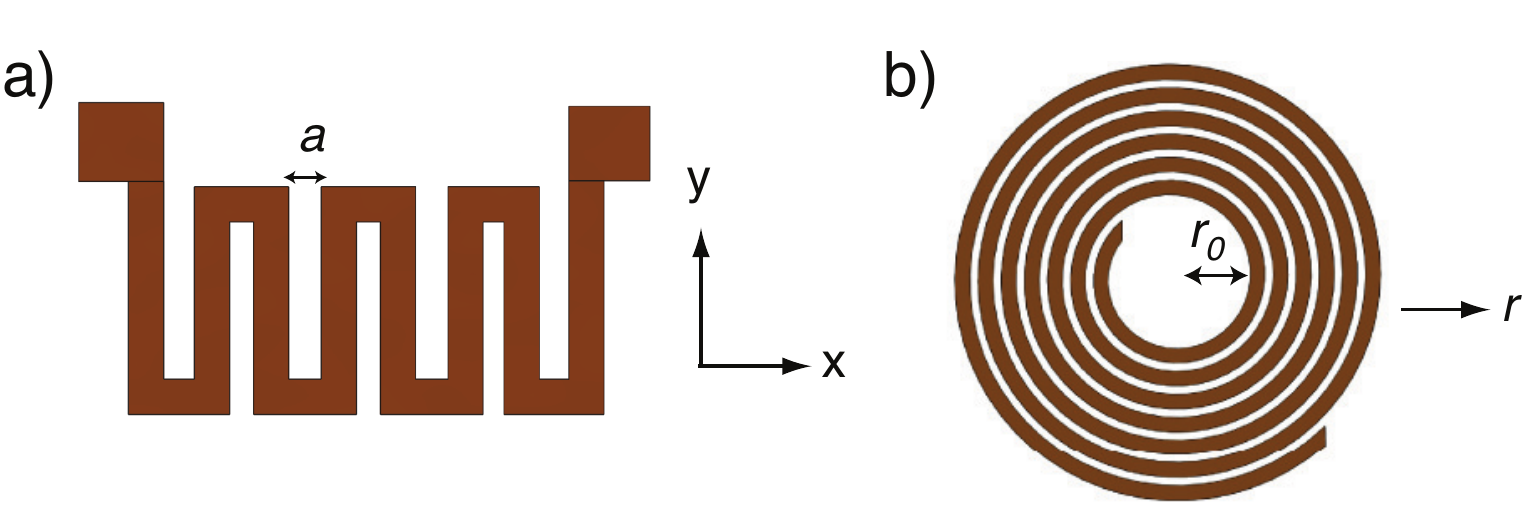}} 
\begin{minipage}{1\hsize}
 \vspace*{-0.1cm}
\caption[]{\label{Fig2SurfCoils}{Sketches of two different geometries of surface coils investigated here: {\bf a)} meander-line and {\bf b)} spiral.
Mutual separation  between the parallel conductors of the meander-line coil is denoted by $a$, while $r_{0}$ denotes the opening radius of the spiral. } 
 }
 \vspace*{-0.1cm}
\end{minipage}
\end{minipage}
\end{figure}
%
%

\section{Theory}
\label{theory}

The optimal choice of  surface coil geometry relies  on the specifics of a particular application. 
We will discuss the two geometries   illustrated   in \mbox{Fig. \ref{Fig2SurfCoils}}. They differ  in uniformity of the in-plane RF fields. 
  These are   meander-line and Archimedean spiral coils. As shown in  \mbox{Fig. \ref{Fig2SurfCoils}a}, the meander-line coil is a serpentine array of parallel conductors of mutual separation $a$. The archimedean spiral is defined as $r = s \theta$ in polar coordinates, where $s$ is the spiral constant.  Coils of spiral geometry  induce more uniform in-plane RF fields, and thus represent a more appropriate geometry for  applications requiring  uniform excitations. 
  
  The excitation field pattern of the surface coil is fully described by the spatial distribution of the magnetic field ($B_1$) produced by   unit current through the coil, and   can be calculated from the Biot-Savart law   \cite{Letch89}.
  For a given surface coil, excitation and reception patterns are identical \cite{BlumichBook}. Therefore by calculating the   spatial distribution of the $B_1$ magnetic field, one can also obtain   information about the  reception sensitivity of the coil. In materials  such as metals,  where RF field penetration is spatially non uniform, calculation of the  spatial distribution of  $B_1$  does not necessarily give precise reception sensitivity. Thus, to compare true reception sensitivity of different coils in our study  we have  independently optimized excitation pulses for each coil and applied field orientation.
  
  The   spatial distribution of the $B_1$ magnetic field has been calculated  for both spiral \cite{Eroglu03} and meander-line \cite{Letch89,Buess91} coils.  In what follows, we give a brief overview of the main results of these calculations. 
  The distribution of the magnitude of  $B_1$ calculated near a conductor of a meander-line coil, defining the $x-y$ plane, in the near field region, for $z/a \ll 1$, is given by 
\begin{equation}
 \label{meanderline_B}
B_1 (x,y,z)= \mu_0 I \frac{e^{-\pi z /a}} {2\pi z (1 + x^2 / z^2 )^{1/2}} \, ,
\end{equation}
for $-a/2 \leqslant x \leqslant a/2$. Here $z$ is the distance away from the coil plane, and the coil strips are considered as infinitely thin but of finite width. The magnitude of $B_{1}$ is the quantity of interest in NQR detection, while the individual components of $B_{1}$ perpendicular to the applied field define excitation/detection pattern in NMR. The important finding is that  the RF magnetic field  has the periodicity of the meander-line itself,  in planes parallel to the surface of a meander-line coil. However, its strength at a distance $z$ away from the coil is given by $\exp(-\pi z/a)$. Therefore, the effective RF field 
  is confined to a region adjacent to the coil and its penetration depth is determined only by the spacing $a$ and not by the overall size of the coil.  Since the signal-to-noise ratio in an NMR experiment is proportional to the filling fraction, the meander-line coils are ideally suited for probing thin 2D-like samples or   sizable surface area to a limited depth \cite{BlumichBook, Buess91}. Consequently, by adjusting only the spacing $a$, one can control SNR and sensitivity depth.

 For spiral coils, the on axis field decays more slowly at a distance $z$ away from the coil plane than that for the meander-line coil. In the limit of spacing between conducting traces $(\Delta)$ going to zero, on-axis field reduces to the  expression for a single loop RF coil carrying a current  $I$ in the AC conductor limit \cite{BlumichBook, Eroglu03},
 \begin{equation}
 \label{meanderline_B}
B_1 (0,z) = {\mu_0 \over 2} I \frac{r_{0}^{2}} {(r_{0}^{2} +  z^2 )^{3/2}} \, .
\end{equation}
Evidently, the RF field and consequently SNR decreases with  increasing axial distance from the coil's center. 
Total $B_{1}$ field can be increased by adding turns to the spiral. Each additional turn strengthens total $B_{1}$ field by superposition.
However, as additional turns add to the total resistance to the coil, there is an optimal number of turns for each coil configuration that maximizes SNR.  For our designs we did not find decreasing reception sensitivity for the coils with up to 7 turns.

\section{Experimental}
\label{exp}
\subsection{NMR set-up} 

The measurements were done using a high homogeneity superconducting magnet with field strength of 7 T. 
Data was taken at room temperature. The NMR data was recorded using a state-of-the-art laboratory-made NMR spectrometer. 
The coils were mounted on a homemade broad band NMR probe, constructed based on the design described in \cite{ArneilProbe}. 
Variable capacitors were used to tune each coil to the desired resonance frequency and assure that circuit is matched to $50 \, \Omega$. 
NMR absorption spectra were obtained from the Fourier transform of the spin-echo.  We used a standard spin echo sequence $(\pi/2-\tau-\pi)$, with pulses independently optimized for each coil and applied field orientation. This was done to assure that only reception sensitivity is compared in each case. Because the samples are highly inhomogeneous, 
only a single line was observed for \In   NMR.  That is, we did not observe nine distinct quadrupolar satellite lines expected for  \In with nuclear spin   $I = 9/2$ in non-cubic local environment.   
 
 \subsection{Coil fabrication}

We used  CST Microwave Studio, a popular commercial tool for 3D EM simulation of microwave and RF frequency, to build 3D models of the various surface coils. This software generates the 3D field distribution using a discretized solution of the integral formulation of Maxwell's equation. The calculated field distribution depends on the exact boundary conditions used. We used open boundary conditions for the spatial part  and in time domain by a so-called ``transient finite difference time domain'' approach. These conditions produce results that are 
sufficiently accurate for MRI imaging applications. However, we used the software just as a rough guide for the field distribution and used our measurements to identify an  ideal coil geometry for a particular application. 
 More importantly, the CST Microwave Studio  output file is used by CircuitCAM software to build the design for machining the coil. 
  CircuitCAM outputs the file in .lmd format (common PCB format) which is then loaded to BoardMaster software to control CNC machine for physical coil fabrication. We used \mbox{300 $\mu$m}  thick  PC board with Cu conducting layer of thickness of \mbox{25 $\mu$m} to machine the coils of desired geometry.  Coils were fabricated at the NMR facility, Division of Structural and Synthetic Biology
Centre for Life Science Technologies, RIKEN Yokohama Campus, Japan, led by Prof. H. Maeda. 

We  also fabricated meander-line micro-coils with  \mbox{100 nm}  spacing between the conductors using lithographic techniques. The gold conducting leads were wire bound and used to connect to the NMR tank circuit. We were able to send sufficient power to detect an NMR echo signal without burning the coil with the RF power. These type of coils can be very beneficial for the NMR studies of nano-devices.

\subsection{Test sample}

For this study, we used \mbox{400 $\mu$m} thick InP substrate as our test samples made at the IBM research center in the group of Dr. D. Sadana. The InP substrates are semi-insulating films doped with Fe, with room temperature resistance exceeding 1 M$\Omega$.  Such samples  were chosen  for the following reasons. 
Both \In and \P nuclei provide good NMR sensitivity. These substrates can be easily cut to the desired shape, allowing   us  
to perform a comprehensive set of desired performance tests. The thickness of the substrates was selected to be comparable to the spacing between neighboring Cu conductors in the surface coil.  For meander-line coils this   assured that the effective RF field would penetrate the entire thickness of the sample. To eliminate artifacts associated with the skin-depth and  induced  surface currents, Fe doped insulating InP samples were taken.

\section{Results and Discussion}
\label{res}
\begin{figure}[t]
\begin{minipage}{0.98\hsize}
\centerline{\includegraphics[scale=0.49]{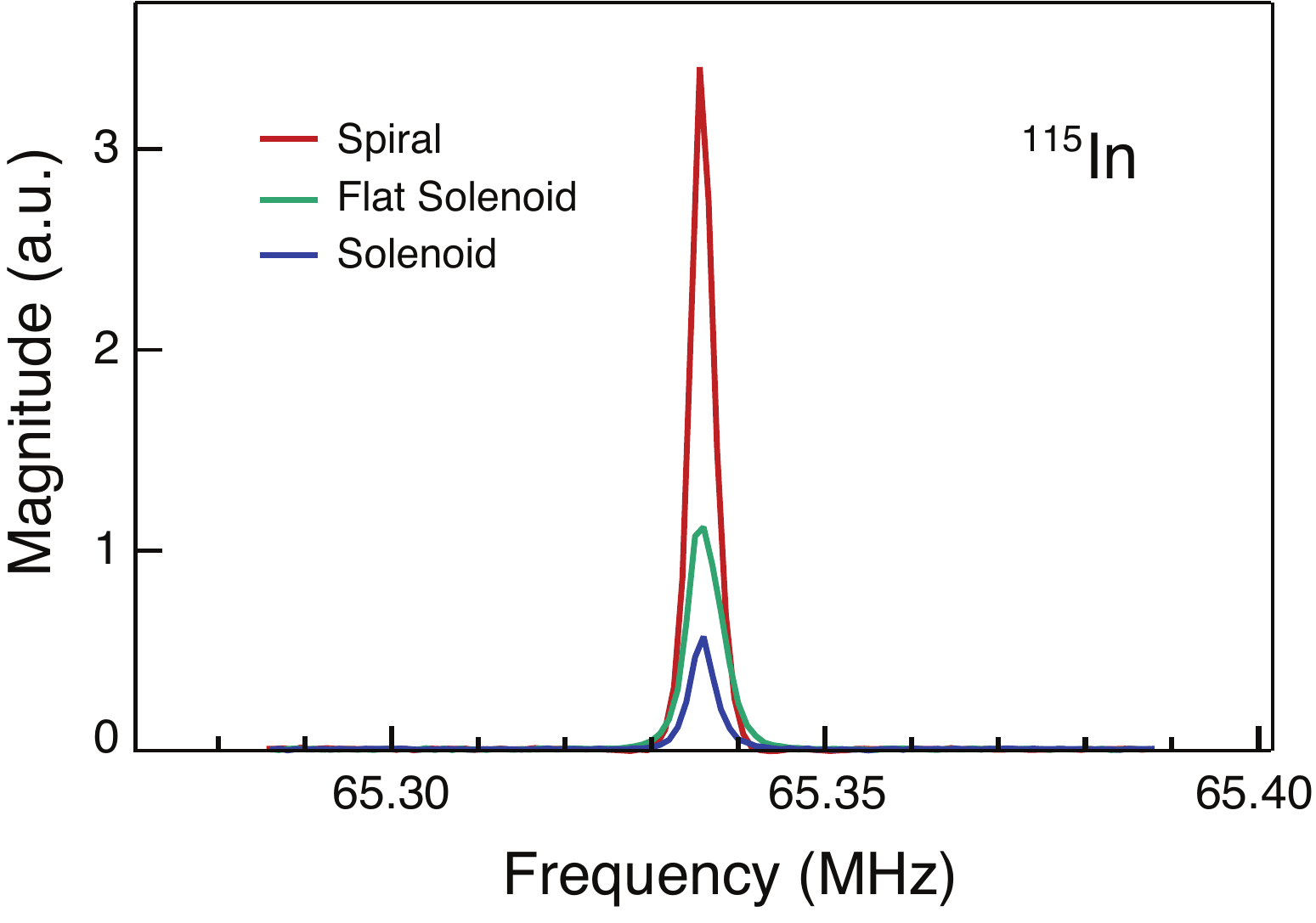}} 
\begin{minipage}{1\hsize}
        \caption[]{\label{Fig3SigComp}{Comparison of the magnitude of the In signal, \ie \In spectra, recorder using three different coils (solenoid, spiral, and flat solenoid) illustrated in \mbox{Fig. \ref{Fig1Coils}}. Square sample of 3 mm $\times$ 3 mm was used.  The applied field was oriented parallel to the plane of the spiral coil. Optimal excitation pulse conditions were determined separately for each coil. Relative ratios of the magnitudes are approximately 6:2:1.}   }
\vspace*{-0.2cm}
\end{minipage}
\end{minipage}
\end{figure}

In a typical NMR experiment a sample is placed in a solenoid coil, as described above. Such coil provides very low filling fraction, and thus low SNR, for flat 2D-like samples. This can be easily resolved in part by using a flat solenoid, as depicted in \mbox{Fig. \ref{Fig1Coils}b}. As a matter of fact, we found that SNR can be improved by a a factor of two, as shown in  \mbox{Fig. \ref{Fig3SigComp}}. However, significant improvement of reception sensitivity by a factor of six  can be achieved when a surface coil is employed as compared to the solenoid coil. In  \mbox{Fig. \ref{Fig3SigComp}}, we plot magnitude of the \In NMR spectra  for the same square sample acquired using three different coils, as denoted. The results clearly demonstrate the advantage of the surface coil. Here, we used the spiral surface coil, with the coil plane 
 oriented parallel to the applied field and the sample covering significant area of the coil. 

Our next step is to determine the optimal geometry of the surface coil for a given sample size and the appropriate variation of the external parameters, such as the applied magnetic field. This is particularly important when specific anisotropic and/or inhomogeneous quantities are to be 
investigated. 
 
%
%
%
%

\subsection{Coil plane parallel to the applied magnetic field}

In this section, we will examine the performance of different surface coils with the plane 
 oriented parallel to the applied field. We first consider spiral coils. As discussed in \mbox{Sec. \ref{theory}}, the strongest effective RF field is induced at the center of the spiral and quickly decays as one moves away from the plane of the coil. 
  When the applied field is oriented parallel to the surface of the coil, this component of the effective RF field is responsible for the spin flip since it is perpendicular to the applied field. 
 This also implies that the most sensitive reception region is in the center cavity of the spiral coil.   
 Therefore, for this applied field orientation, we expect that placing the sample in the center cavity of the spiral should provide the best reception sensitivity, and consequently SNR.

%
\begin{figure}[t]
\begin{minipage}{0.98\hsize}
\centerline{\includegraphics[scale=0.45]{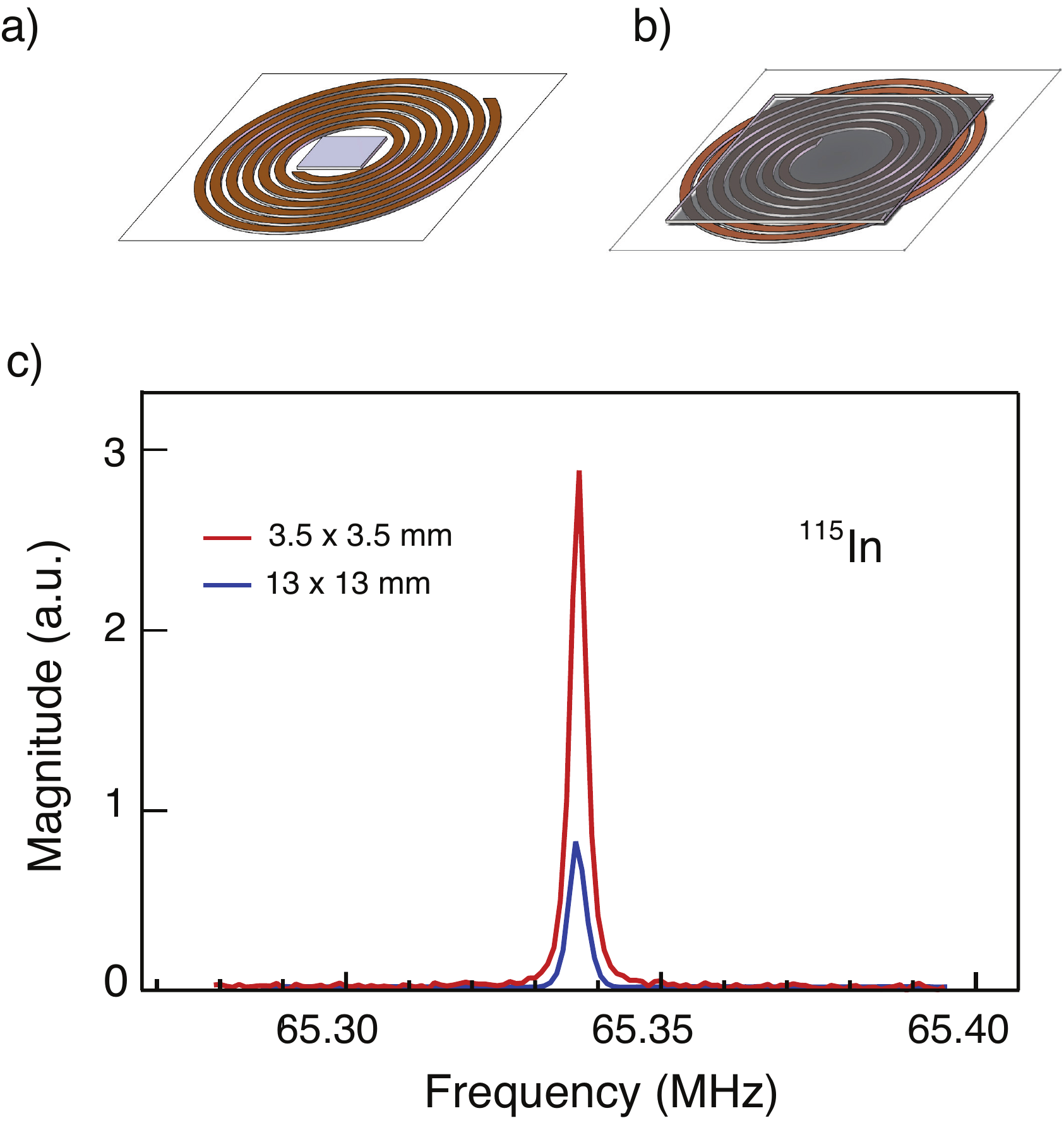}} 
\begin{minipage}{1\hsize}
\caption[]{\label{Fig4Spir}{Sketch of the two samples of different sizes and their placement in the plane of the spiral surface coil: {\bf a)}  \mbox{3.5 $\times$ 3.5 mm} in the cavity of the spiral coil and {\bf b)} \mbox{13 $\times$ 13 mm} covering most of the area of the surface coil.   
{\bf c)} Comparison of the magnitude of the \In signal per unit area from two samples, placed  in the plane of the spiral surface coil as depicted in part  {\bf a)}  and  {\bf b)}.  The normalized signal per unit area from the sample placed in the cavity of the spiral surface coil  is        approximately 3.7 times that of the sample covering the entire area of the coil. }   }
\vspace*{-0.2cm}
\end{minipage}
\end{minipage}
\end{figure}
%

To test this hypothesis, we used the same spiral coil to obtain NMR signal from two samples of different sizes, as illustrated in \mbox{Figs. \ref{Fig4Spir}a \& b}. One sample is such that it can be  entirely placed  into the center cavity of the spiral, while the other covers most of the area of the same coil. 
  In  \mbox{Fig. \ref{Fig4Spir}c}, we plot magnitude of the \In NMR signal per nucleus acquired by the same spiral coil  for  samples of different sizes, placed as shown in the figure. For these particular geometries, we find that the magnitude of the signal for the case when the sample is in the cavity exceeds by more then a factor of three that when the sample covers the coil. 
  This result confirms the hypothesis, that highest sensitivity can be obtained by placing the sample in the center cavity of the spiral.

Therefore, for this applied field orientation and given sample size the best sensitivity can be achieved by designing the spiral coils such that the entire sample can be placed into the center cavity. The sensitivity can be further improved by increasing the  effective RF field  by adding turns to the spiral. The requirement that the tank circuit must tune to resonance frequency of a particular nuclear species  imposes an upper bound on the total number of  coil turns. Therefore,  high frequency applications of the spiral coil with the sample placed in the center cavity  might not offer significant gain in sensitivity as compared to other coil geometries  such as meander-line. 
 
%
\begin{figure}[b]
\begin{minipage}{0.98\hsize}
\centerline{\includegraphics[scale=0.52]{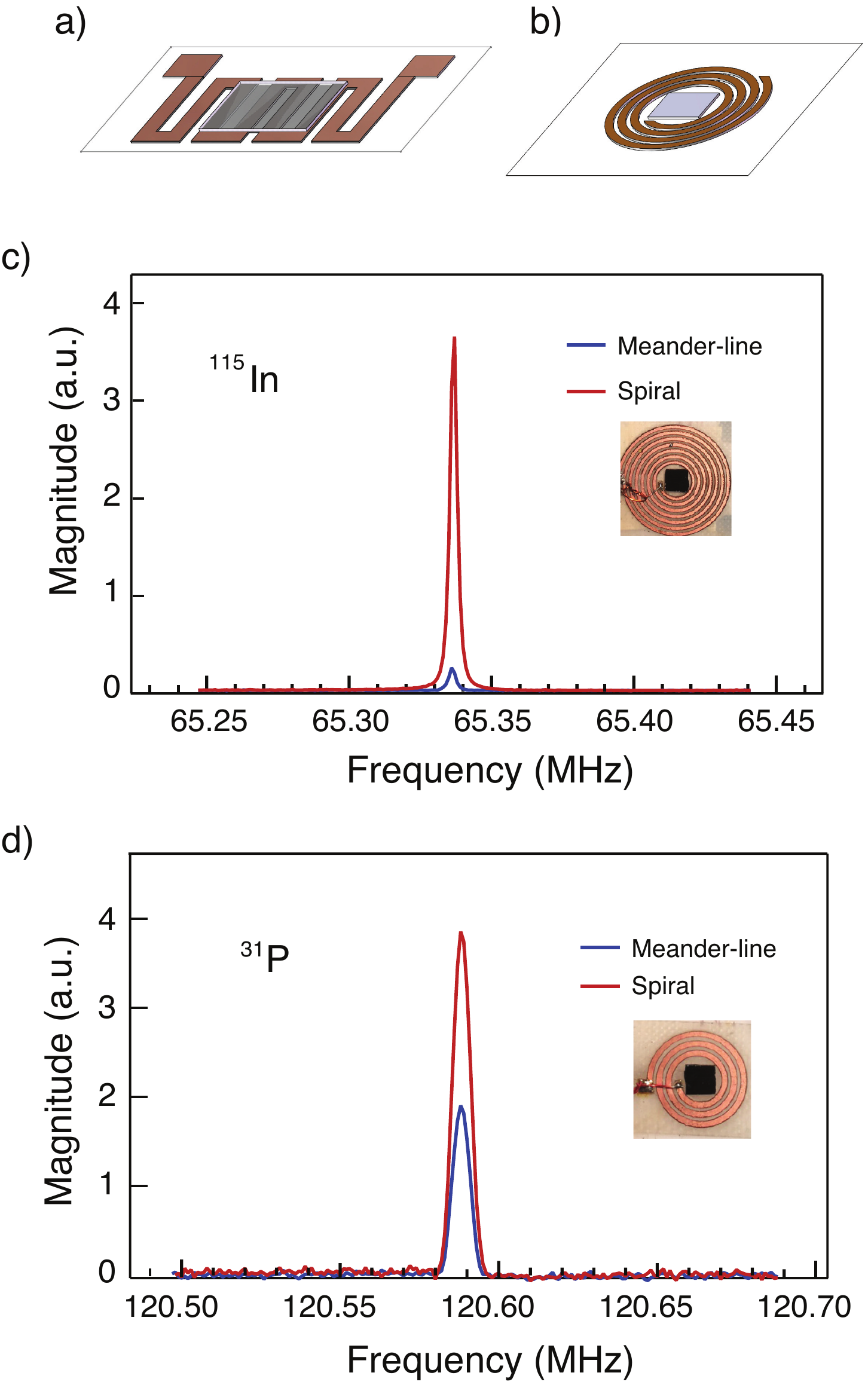}} 
\begin{minipage}{1\hsize}
\caption[]{\label{Fig5}{ 
Sketch of the surface coils and  placement of the sample  in the plane of the  meander-line   {\bf a)}  and spiral coil  {\bf b)} for measurements plotted in part  {\bf c)} and  {\bf d)}.  
Sample of \mbox{3.5 $\times$ 3.5 mm} area is used. Diameter of central cavity of both spiral coils is \mbox{$\approx 5$ mm}, \ie opening radius of the spiral is \mbox{$r_{0} = 2.5$ mm}. The same meander-line coil is used in both cases. 
{\bf c)} Relative magnitude of the \In signals obtained using meander-line and spiral surface coils. Signal detected by dense 7-turns spiral is nearly 16 times stronger.  Diameter of the  coil is 17.1 mm with conductor width of \mbox{510 $\mu {\rm m}$} {\bf d)} Relative magnitude of the \P signals obtained using meander-line and spiral surface coils. Signal detected by 3-turns spiral, necessary to tune to higher frequencies, exceed that obtained by the meander-line coil by a factor of 2. Diameter of the  spiral coil is 11.1 mm with conductor width of \mbox{850 $\mu {\rm m}$}}.   }
\vspace*{-0.2cm}
\end{minipage}
\end{minipage}
\end{figure}
%

Next, we compare the signal magnitude obtained using meander-line and spiral surface coils, as depicted in \mbox{Fig. \ref{Fig5}a \& b}. The comparison is performed at two separate frequencies,  differing by nearly a factor of two, corresponding to \In and \P resonances. 
Two spiral coils with the same size of the center cavity are employed to achieve the required resonance frequencies. In both cases, the sample is positioned in the center cavity of the coil. The spiral with less turns is used for higher frequency \P NMR, as displayed in \mbox{Fig. \ref{Fig5}c \& d}. 

 In  \mbox{Fig. \ref{Fig5}c}, we plot the magnitude of  \In NMR signals acquired by different surface coils. Two traces are acquired  from the same sample, as depicted. One from the sample being placed  in the cavity of the spiral coil and  the other on the meander-line.  The signal detected by   a dense seven turns spiral is nearly 16 times stronger then that detected by the meander-line coil. As previously discussed, decreasing the number of turns of the spiral lowers the effective RF field in the center, and thus the sensitivity. We demonstrate this effect  by comparing the signal  detected from the same sample by   a three turn spiral and  the meander-line.  As shown in  \mbox{Fig. \ref{Fig5}d}, signal detected by   a three turns spiral is  only 2 times stronger then that detected by the meander-line coil. Evidently, signal detected by the dense 7-turns spiral is nearly 8 times that detected by that with three turns.

\subsection{Coil Plane perpendicular to the applied magnetic field}

In what follows, we will examine the performance of surface coils when their plane is oriented perpendicular to the applied field. 
This is an  important case, as this geometry is required in the study of physical phenomena such as vortices in superconductors and quantization of 2D electron gas. In the previous section, we established that placing the sample in the center cavity of the spiral coil  provides the best reception sensitivity  because the strongest effective RF field is induced at the center of the spiral. However,  for the applied field oriented perpendicular to the plane of the coil, this RF field,  being   aligned with the applied field,  cannot induce spin flips. Therefore, no NMR signal can be detected in this geometry.

%
\begin{figure}[h]
 \vspace*{0.1cm}
\begin{minipage}{0.98\hsize}
\centerline{\includegraphics[scale=0.37]{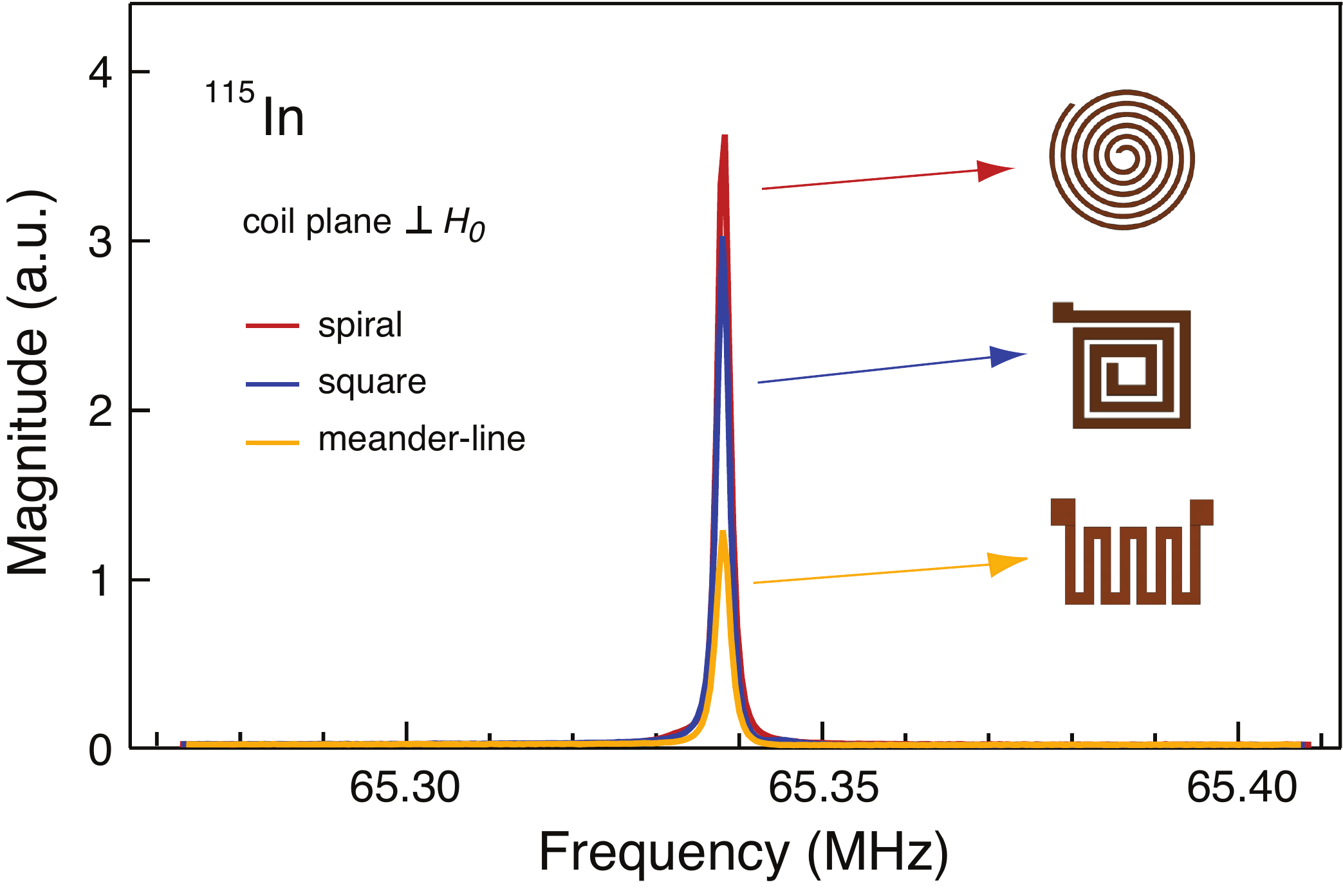}} 
\begin{minipage}{1\hsize}
\caption[]{\label{Fig6}{Relative magnitude of the \In signals obtained using meander-line, square, and spiral surface coils with the plane of the coil oriented perpendicular to the applied field, $H_{0}$. Sample of \mbox{3.5 $\times$ 3.5 mm} area is used. 
Relative magnitude of signals scale as 6:5:2, for  spiral, square, and meander-line coils respectively. We used coils with the following  specifications: square coil of  10.1 mm in length, with the   opening radius/length of \mbox{$r_{0} = 1.05$ mm}, and \mbox{1060 $\mu {\rm m}$} conductor width;  spiral coil of  14 mm in diameter, with the   opening radius of the spiral of \mbox{$r_{0} = 1.1$ mm}, and \mbox{580 $\mu {\rm m}$} conductor width; and,   meander-line coil of  10.2 mm in length and 3.4 mm width, with spacing between conducting lines of \mbox{$\approx 1$ mm},  and \mbox{530 $\mu {\rm m}$} conductor width.  }   }
\vspace*{-0.2cm}
\end{minipage}
\end{minipage}
\end{figure}

For this applied field orientation, the  NMR signal is generated by the effective in-plane RF fields. 
We  compare   the NMR signal acquired on the same sample by three different coil geometries that produce such RF fields. 
 In  \mbox{Fig. \ref{Fig6}}, we plot the magnitude of  \In NMR signals acquired by meander-line, square, and spiral surface coils, as denoted. In all three cases the sample covers a significant area of the coil. For this insulating sample, the signal acquired by  the meander-line coil is the weakest. However,  for studies of metallic samples with finite skin depth the use of a meander-line coil with a spacing between conducting lines that matches the skin depth provides  optimal filling fraction, and thus sensitivity, as described in \mbox{Sec. \ref{theory}}.
 Another example of potentially effective use of meander-line coils includes NMR studies 
 of vortices in superconductors. Here,  the optimal sensitivity is achieved by a spacing between conducting lines that matches the  superconducting penetration depth.

\subsection{Different orientations of  the applied magnetic field}

To study intrinsic anisotropies, it is crucial to investigate sample properties as a function of the orientation of the applied magnetic field with all other parameters being fixed. It important to compare relative magnitude of the signals acquired with the plane of the coil oriented parallel and perpendicular to the applied field. Next, we present such comparison for three different coil designs described above.

%
\begin{figure}[h]
\begin{minipage}{0.98\hsize}
\centerline{\includegraphics[scale=0.41]{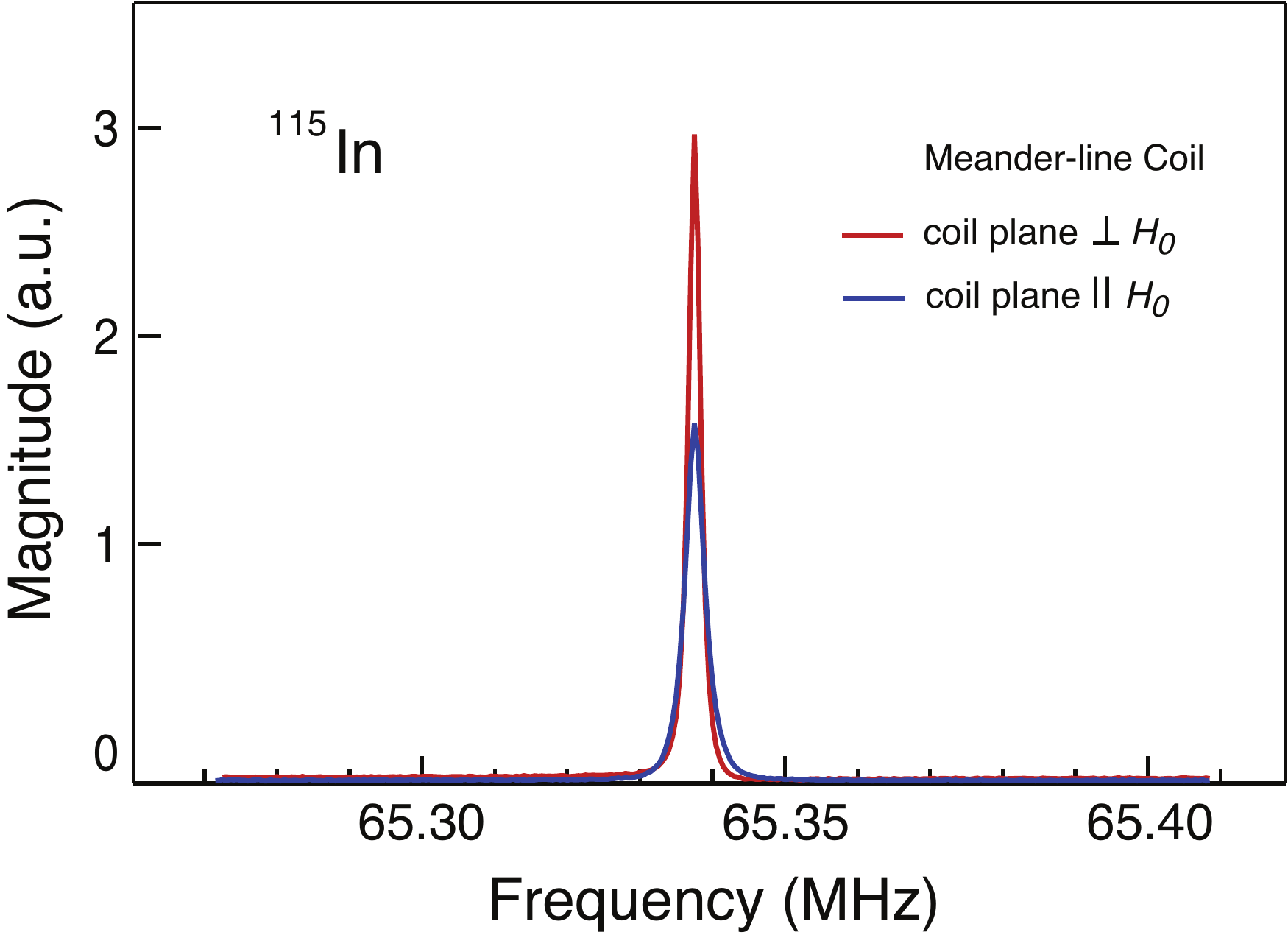}} 
\begin{minipage}{1\hsize}
\caption[]{\label{Fig7}{Relative magnitude of the \In signals obtained using meander-line  surface coil with the plane of the coil oriented parallel and perpendicular to the applied field, $H_{0}$. When the plane of the coil is perpendicular to the applied field, a signal 1.8 times stronger is detected.  We use meander-line coil of  10.2 mm in length and 3.4 mm width, with spacing between conducting lines of $\approx 1$ mm,  and \mbox{530 $\mu {\rm m}$} conductor width. }   }
\vspace*{-0.2cm}
\end{minipage}
\end{minipage}
\end{figure}

%
\begin{figure}[h]
\begin{minipage}{0.98\hsize}
\centerline{\includegraphics[scale=0.35]{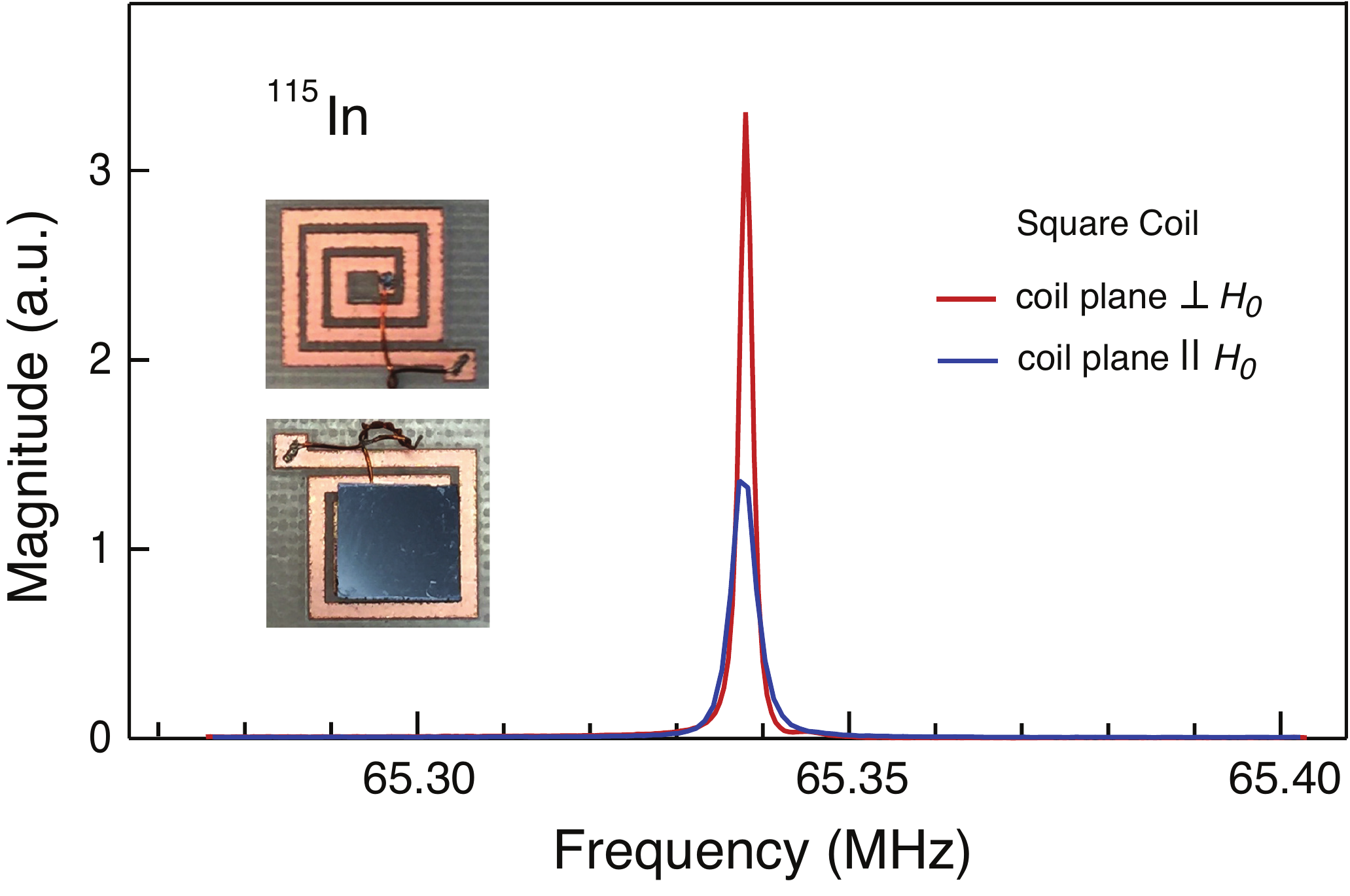}} 
\begin{minipage}{1\hsize}
\caption[]{\label{Fig8}{Relative magnitude of the \In signals obtained using square surface coil with the plane of the coil oriented parallel and perpendicular to the applied field, $H_{0}$. When the plane of the coil is perpendicular to the applied field, a signal 2.5 times stronger  is detected. Photographs depict the coil and the sample, black square, placement. We use square coil of  10.1 mm in length, with the   opening radius/length of \mbox{$r_{0} = 1.05$ mm}, and \mbox{1060 $\mu {\rm m}$} conductor width.   }   }
\vspace*{-0.2cm}
\end{minipage}
\end{minipage}
\end{figure}

\begin{figure}[h]
\begin{minipage}{0.98\hsize}
\centerline{\includegraphics[scale=0.43]{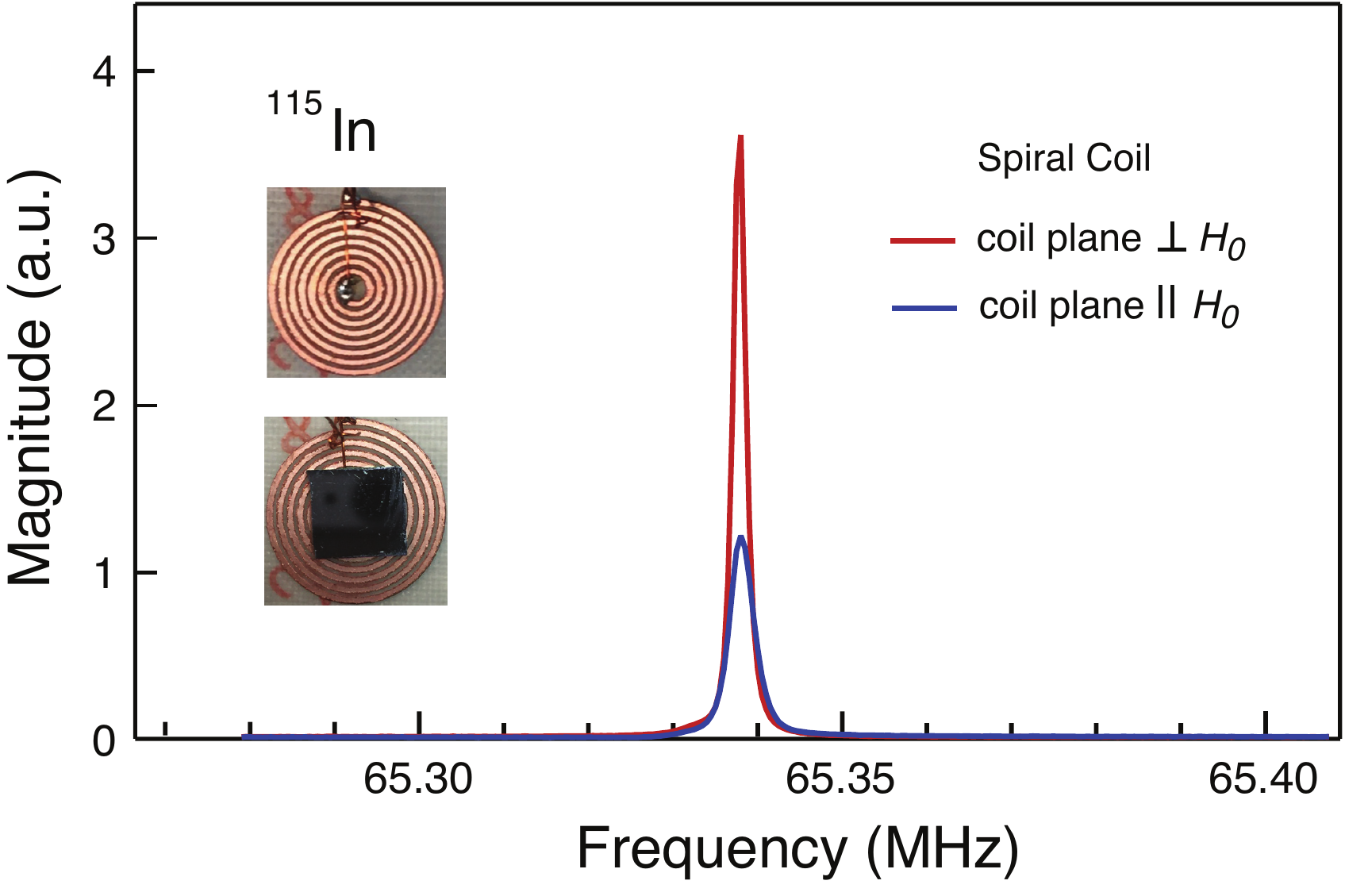}} 
\begin{minipage}{1\hsize}
\caption[]{\label{Fig9}{Relative magnitude of the \In signals obtained using spiral surface coil with the plane of the coil oriented parallel and perpendicular to the applied field, $H_{0}$. When the plane of the coil is perpendicular to the applied field, a signal 3 times stronger is detected.
Photographs depict the coil and the sample, black square, placement. We use spiral coil of  14 mm in diameter, with the   opening radius of the spiral of \mbox{$r_{0} = 1.1$ mm}, and \mbox{580 $\mu {\rm m}$} conductor width.  
}   }
\vspace*{-0.2cm}
\end{minipage}
\end{minipage}
\end{figure}

Relative magnitude of the \In signals obtained using meander-line, square, and spiral  surface coil with the plane of the coil oriented parallel and perpendicular to the applied field is plotted in  \mbox{Fig. \ref{Fig7},  \ref{Fig8}, and  \ref{Fig9}}, respectively. 
The smallest  variation in detected magnitude for two field orientations is observed for signal acquired by a meander-line coil. 
That is, when the plane of meander-line coil is perpendicular to the applied field, a signal 1.8 times stronger  is detected.  
On the other hand,  the signal for a spiral coil  varies by a factor of three for two different field orientations. 
Therefore, meander-line surface coils are best suited for the studies of  intrinsic anisotropic properties as their use minimizes artifacts associated with the effective RF field anisotropy.  

%

\section{Conclusions}

We investigate the performance of surface coils for NMR studies of quasi 2D materials. That is, we compared the reception sensitivity of surface micro-coils of  spiral and meander-line  geometries. The optimal geometry of the surface coil for a given application and the direction of the applied field is identified. In this study we were not concerned with the homogeneity of the effective RF field induced by the coil, since the phenomena in condensed matter systems often produce intrinsic inhomogeneity exceeding by far those associated by the RF field.   

For the magnetic field applied parallel to the coil plane, the best sensitivity can be achieved by employing the spiral coil designed so that the entire sample can be placed in the center cavity of the coil. However, this geometry will yield no signal for the magnetic field applied perpendicular to the coil plane. In this case, the best sensitivity can be achieved by employing the spiral coil designed so that the  sample covers a significant area of the coil.    

Meander-line surface coils are best suited for the studies of  intrinsic anisotropic properties, since  their use minimizes artifacts associated with the effective RF field anisotropy.  Furthermore, meander-line geometry is best suited for the study of samples in which RF penetration is spatially inhomogeneous, such as metals and superconductors. In this case, optimal sensitivity is obtained if the spacing between conducting lines matches the RF penetration depth. Finally, in addition to  the gain in sensitivity for 2D-like samples, application of the surface coils offers ready access to the sample, which can be crucial in allowing {\it in-situ} variation of parameters, such as bias gate voltage.

\begin{acknowledgments}
 We would like to thank  Dr. Devendra K, Sadana, IBM - T. J. Watson Research Center,    for  providing InP thin substrate samples for our work, 
 and Prof. Hideaki Maeda,  RIKEN Yokohama Campus, for hosting Lu Lu and making coil fabrication facility available to us. We acknowledge  guidance for software use from Dr. Yoshinori Yanagisawa and Dr. Masato Takahashi.  
     This research was supported in part by   the National Science Foundation under Grant No. DMR-1608760.
     \end{acknowledgments}

  \bibliographystyle{ieeetr}
\bibliography{SurfCoil_Ref}

\begin{thebibliography}{10}

\bibitem{AbragamBook}
A.~Abragam, {\em Principles of Nuclear Magnetism}.
\newblock Oxford University Press, 1985.

\bibitem{Buess91}
M.~L. Buess, A.~N. Garroway, and J.~B. Miller, ``{NQR} detection using a
  meanderline surface coil,'' {\em J. Mag. Res.}, vol.~92, p.~348, 1991.

\bibitem{Eroglu03}
S.~Eroglu, B.~Gimi, B.~Roman, G.~Friedman, and R.~L. Magin, ``{NMR spiral
  surface microcoils: Design, fabrication, and imaging},'' {\em Concepts in
  Magnetic Resonance Part B: Magnetic Resonance Engineering}, vol.~17B, no.~1,
  pp.~1--10, 2003.

\bibitem{Ackerman80}
J.~J. Ackerman, T.~H. Grove, G.~G. Wong, D.~G. Gadian, and G.~K. Radda,
  ``{Mapping of metabolites in whole animals by $^{31}$P NMR using surface
  coils},'' {\em Nature}, vol.~283, pp.~167--170, January 1980.

\bibitem{BlumichBook}
B.~Bl{\"u}mich, {\em NMR Imaging of Materials}.
\newblock Oxford University Press, 2000.

\bibitem{Lu17}
L.~Lu, M.~Song, W.~Liu, A.~P. Reyes, P.~Kuhns, H.~O. Lee, I.~R. Fisher, and
  V.~F. Mitrovi{\'c}, ``{Magnetism and local symmetry breaking in a Mott
  insulator with strong spin orbit interactions},'' {\em Nature
  Communications}, vol.~8, p.~14407 EP, 02 2017.

\bibitem{MitrovicLSCO}
V.~F. Mitrovi{\'c}, M.-H. Julien, C.~de~Vaulx,
  M.~Horvati\ifmmode~\acute{c}\else \'{c}\fi{}, C.~Berthier, T.~Suzuki, and
  K.~Yamada, ``{Similar glassy features in the $^{139}$La NMR response of pure
  and disordered La$_{1.88}$Sr$_{0.12}$CuO$_{4}$},'' {\em Phys. Rev. B},
  vol.~78, p.~014504, Jul 2008.

\bibitem{Koutroulakis08}
G.~Koutroulakis, V.~F. Mitrovi\ifmmode~\acute{c}\else \'{c}\fi{},
  M.~Horvati\ifmmode~\acute{c}\else \'{c}\fi{}, C.~Berthier, G.~Lapertot, and
  J.~Flouquet, ``{Field Dependence of the Ground State in the Exotic
  Superconductor ${\mathrm{CeCoIn}}_{5}$: A Nuclear Magnetic Resonance
  Investigation},'' {\em Phys. Rev. Lett.}, vol.~101, p.~047004, Jul 2008.

\bibitem{Letch89}
J.~H. Letcher, ``Computer-assisted design of surface coils used in magnetic
  resonance imaging. {I}. {T}he calculation of the magnetic field,'' {\em Mag.
  Res. Imag.}, vol.~7, p.~581, 1989.

\bibitem{ArneilProbe}
A.~P. Reyes, H.~N. Bachman, and W.~P. Halperin, ``{Versatile 4 K nuclear
  magnetic resonance probe and cryogenic system for small-bore high-field
  Bitter magnets},'' {\em Review of Scientific Instruments}, vol.~68, no.~5,
  pp.~2132--2137, 1997.

\end{thebibliography}

\end{document}